\documentclass[iop,apjl,twocolumn]{emulateapj_pab}

\usepackage{color}
\usepackage{epstopdf}
\usepackage{graphicx}	
\usepackage{amsmath}	
\usepackage{amssymb}	
\usepackage{makeidx}
\usepackage{listings}
\usepackage{array}
\usepackage{times}
\usepackage{rotating}
    
\graphicspath{{./}{Bilder/}}

\renewcommand*\aap{A\&A}

\renewcommand*\aj{AJ}
\renewcommand*\ao{Appl.~Opt.}

\renewcommand*\apj{ApJ}
\renewcommand*\apjl{ApJL}

\renewcommand*\apjs{ApJS}

\renewcommand*\araa{ARA\&A}

\renewcommand*\mnras{MNRAS}

\renewcommand*\nat{Nature}

\renewcommand*\pasp{PASP}

\renewcommand*\qjras{QJRAS}

\renewcommand{\AA}{\r{A}}

\definecolor{darkgreen}{rgb}{0,0.45,0}

\begin{document}
\hypersetup{
	pdftitle = {SDSS J1056+5516: A Triple AGN or an SMBH Recoil Candidate?},
	pdfauthor = {E.~Kalfoutnzou},
	pdfkeywords = { galaxies: active -- galaxies: evolution -- galaxies: interactions},
	pdfsubject = {The Astrophysical Journal Letters, 851:L51 (6pp), 2017 December 10}
}


\title{{\bf SDSS J1056+5516: A Triple AGN or an SMBH Recoil Candidate?}}
\shorttitle{Triple AGN or SMBH recoil?}
\shortauthors{Kalfountzou}

\author{E.~Kalfountzou\altaffilmark{1}, M Santos Lleo\altaffilmark{1} and M. Trichas\altaffilmark{2}}
\affil{$^{1}$European Space Astronomy Center (ESA/ESAC), Mission Operations Division, P.O. Box 78, E-28691 Villanueva de la Ca$\tilde{n}$ada, Madrid, Spain; \href{mailto:ekalfountzou@sciops.esa.int}{ekalfountzou@sciops.esa.int}}
\affil{$^{2}$Airbus Defence \& Space, Gunnels Wood Road, Stevenage, Hertfordshire SG1 2AS, UK }

\received{2017 October 18}
\revised{2017 November 6}
\accepted{2017 November 14}
\published{2017 December 7}

\submitted{}
\journalinfo{The Astrophysical Journal Letters, {\rm 851:L15 (6pp), 2017 December 10 \hfill \url{ https://doi.org/10.3847/2041-8213/aa9b2d}}}

\begin{abstract}
We report the discovery of a kiloparsec-scale triple supermassive black hole system at $z=0.256$: SDSS J1056+5516, discovered by our systematic search for binary quasars. The system contains three strong emission-line nuclei, which are offset by $<250~{\rm km~s^{-1}}$ and by 15-18 kpc in projected separation, suggesting that the nuclei belong to the same physical structure. The system includes a tidal arm feature spanning $\sim100$~kpc in projected distance at the systems' redshift, inhabiting an ongoing or recent galaxy merger. Based on our results, such a structure can only satisfy one of the three scenarios; a triple supermasive black hole (SMBH) interacting system, a triple AGN, or a recoiling SMBH. Each of these scenarios is unique for our understanding of the hierarchical growth of galaxies, AGN triggering, and gravitational waves.
\end{abstract}

\section{Introduction} 

In the current $\Lambda$CDM structure formation model \citep[e.g.,][]{Freedman2003}, hierarchical galaxy mergers and interactions play a primary role in galaxy formation \citep[e.g.,][]{Hopkins2005}, with the most massive galaxies expected to harbor a central supermasive black hole (SMBH). In the context of hierarchical merger models, binary SMBHs form frequently, and should be common in galaxies \citep[e.g.,][]{Volonteri2003}. In the cases where the binary lifetime is long enough to exceed the typical time between mergers, then another merger with a third galaxy is possible and some galactic nuclei should contain systems of three or more SMBHs.

These systems are particularly interesting as they produce a range of phenomena rather different from those expected from single SMBHs such as: (1) the formation of elliptical galaxies \citep[e.g.,][]{Hoffman2007}, (2) the trigger of quasars' activity and major SMBH growth \citep[e.g.,][]{Hopkins2005}, (3) the high-velocity ejection of one of the SMBHs (e.g., $\sim10^{3}~{\rm km~s^{-1}}$; \citealp{Komossa2012}), and (4) intense bursts of gravitational radiation \citep[e.g.,][]{Hoffman2007}, which could potentially be observable by forthcoming missions such as pulsar timing arrays (PTAs) and Laser Interferometer Space Antenna (LISA).

Despite their importance, observational evidence for these systems is very limited. The largest catalog of multiple-mergers consists of 39 systems \citep{Darg2011}. Even smaller is the number of triple-AGN candidates, with only 8 at $<100$~kpc separations reported so far (see \citealp{Deane2014} for a review). Similarly, despite the intense search for gravitational-wave recoiling SMBHs, only a few candidates have been proposed (see, e.g., \citealp{Komossa2012}). 
    
In this Letter, we report the discovery of a kiloparsec-scale triple of SMBHs, with a separation large enough that the system can be resolved into its component parts using current facilities, yet small enough to be dynamically interesting. A $\Lambda$CDM cosmology with $\Omega_m = 0.27$, $\Omega_\Lambda =  0.73$, and $h = 0.7$ is assumed throughout \citep{Hinshaw2009}. All the quoted magnitudes are expressed in the AB standard photometric system.
    
\vspace{2cm}    
\section{Data and Analysis} 

\subsection{Discovery and basic properties}

\begin{figure}
\includegraphics[trim=120 40 267 0, clip,scale=0.250]{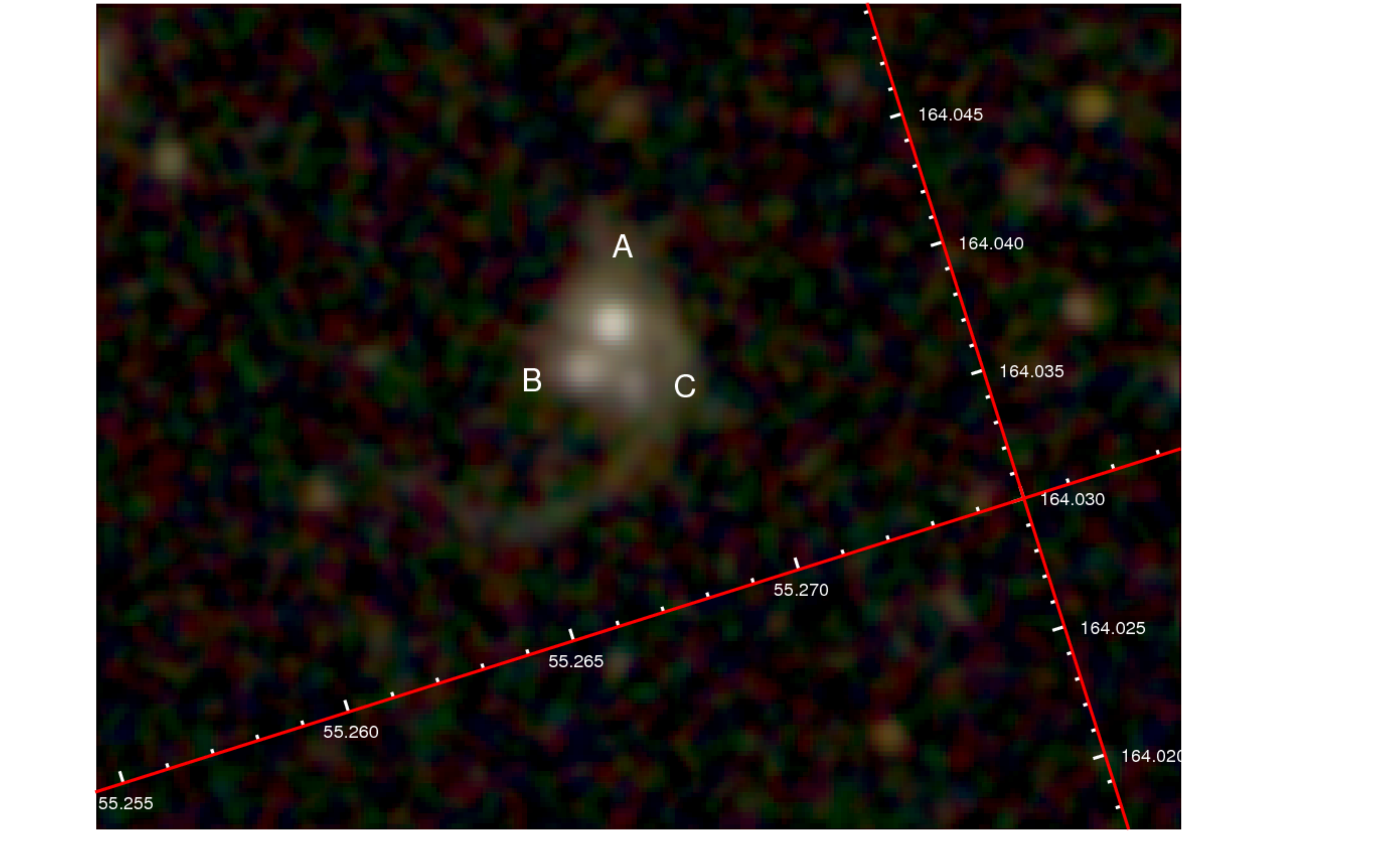}
\caption{SDSS gri--color composite image of SDSS J1056+5516. The system has three nuclei A, B, and C, and a strong tidal feature to the south of C.}
\label{fig:rgi_sdss}
\end{figure}

In the course of searching the Sloan Digital Sky Survey (SDSS) archives for binary quasars, we came across the objects SDSS J105609.79+551604.0 and SDSS J105609.48+551600.9 (hereafter A and B). The first one is found in the SDSS Data Release 7 Quasar catalog (DR7Q; \citealp{Schneider2010}) and the second one is in the SDSS DR13 \citep{Albareti2016}. This system initially appeared to be an excellent example of a binary quasar with a radial velocity separation $\Delta \upsilon_{\rm A,B}={\rm -246~km~s^{-1}}$ and projected distance $\sim16.0$ kpc. Examining the SDSS image of the system, we found a third companion SDSS J105609.21+551604.2 (hereafter C) at the same redshift and projected distance $<20.0$ kpc. Shown in Figure~\ref{fig:rgi_sdss}, the system contains three nuclei, as seen in its SDSS image, along with an extended tidal tail spanning some 100 kpc. Distributed morphologies and/or peculiar features such as tidal tails are clear morphological evidence of interactions or mergers. The details of the basic properties are presented in Table~\ref{table_full}. 

\subsection{Photometric Data}

We have searched the X-ray archives for observations of the system. The unresolved source was detected in the ROSAT all-sky survey source catalog (2RXS; \citealp{Boller2016}) with an X-ray luminosity of $L_{0.1-2.4{\rm keV}} \approx 6\times10^{43}~{\rm erg~s^{-1}}$ corrected for absorption within the Galaxy, using the \cite{Schlegel1998} map and assuming a power-law spectrum with an energy index $\alpha_{\rm X}=1.65$ \citep[e.g.,][]{Sazonov2008}.
    
The SDSS provides images of the system in different filters, and photometry and spectroscopy for each of the three nuclei. Only A is detected by the Two Micron All Sky Survey (2MASS; \citealp{Skrutskie2006}), while the A and B components are blended in the {\it Wide-field Infrared Survey Explorer} ({\it WISE}; \citealp{Wright2010}), and C is not detected. 
    
The B nuclei is the only one detected by the Faint Images of the Radio Sky at Twenty cm survey (FIRST; \citealp{Becker1995}), with $S_{1.4{\rm GHz}} = 1.14 \pm 0.15~{\rm mJy}$ which corresponds to a radio luminosity $\sim 3\times10^{30}~{\rm erg~s^{-1}~Hz^{-1}}$. NRAO VLA Sky Survey (NVSS; \citealp{Condon1998}) observations show a similar level of radio emission suggesting a generally fairly compact nature. Combining optical with radio fluxes, we can constrain the radio-loudness to $R_{i}=\log(F_{\rm radio}/F_i)=1.1$. A radio-loudness parameter greater than one is typically the criterion for a radio-loud classification \citep[e.g.,][]{Ivezic2002}\footnote{Alternatively, the radio emission could be associated with star-formation at the host of $B$ at the level of $\sim 10~{\rm M_{\odot}~yr^{-1}}$.}. The photometric details are presented in Table~\ref{table_full}.
    
\begin{table*}
\addtolength{\tabcolsep}{-3pt}  
 \caption{Properties of the Triple System. For each component we list:  position (R.A., decl.), redshift, angular ($\Delta \theta$; arcseconds) and projected separations ($\Delta \upsilon_{\rm proj}$; kpc), and radial velocity difference ($\Delta V$; ${\rm km~s^{-1}}$), photometric data, SED-fitting outputs, X-ray luminosity, spectroscopic measurements (Flux [${\rm erg~cm^{-2}~s^{-1}}$], FWHM [${\rm km~s^{-1}}$], shift [${\rm km~s^{-1}}$]) .}
\begin{tabular}{l c c c}

    
    %
       &   &   &   \\
    Parameters  & SDSS J1056+5516A & SDSS J1056+5516B & SDSS J1056+5516C \\
           &   &   &   \\
    \tableline
    R.A., Decl.  & 164.0408, 55.2678   & 164.0395,	55.2669   &  164.0384,	55.2678 \\
    redshift & 0.2564  & 0.2572  & 0.2566   \\
    
    $\Delta \theta$ &  4.0 [A,B]  & 3.9 [B,C] & 4.5 [C,A]  \\
    $pd$  & 15.9 [A.B]  & 15.5 [B,C]  &  17.9 [C,A]  \\
    $\Delta V$ & -240.0 [A,B]  & 180.0 [B,C]  &  60.0 [C,A] \\
      &  & &  \\
    Photometry &   &  &  \\

    SDSS ($u,g,r,i,z$)(mag) & [18.70, 18.31, 17.80, 17.59, 17.26] &  [20.64, 19.65, 19.23, 18.93, 18.24]  & [20.26, 20.24, 19.95, 19.10, 20.58]    \\
    MASS2 ($J, H, K$)(mag) & [16.31, 15.23, 14.18]  &  \nodata  &   \nodata \\
    {\it WISE} ($W1,W2,W3,$ & [15.33, 14.91, 13.80, 12.50]  & [15.33, 14.91, 13.80, 12.50]  &   \\
    $W4)$(mag) &   &  &  \\
    FIRST 1.4 GHz(mJy) &  \nodata &  1.14 &  \nodata \\
      &  & &  \\
    
    SED &  &  &  \\
    Stellar Mass   &  \nodata &  \nodata &   $1.5\pm0.4$ \\
     ($10^{10}~M_\odot$)     &  & &  \\

    Stellar Age (Gyr)  &  \nodata &  \nodata  &   $3.2\pm1.1$ \\
    Star Formation   &  \nodata  &   \nodata &   $7.2\pm0.9$ \\
    Rate ($M_\odot/{\rm yr}$) &   &  &  \\

      &   &   &  \\
    $L_{\rm 0.5-2~keV}$ (${\rm erg~s^{-1}}$) & $6.9^{11.3}_{-3.7}\times10^{43}$\footnote{We estimate the X-ray luminosity assuming a ${\rm X/O} = \log(f_{\rm 0.5-2~keV} / f_{i})=0.14^{0.52}_{-0.49}$ \citep{Civano2012} and the relation $\log(f_{\rm 0.5-2~keV} / f_{i}) = f_{\rm 0.5-2~keV} + C_i + 0.4\times m_{i}$ where $C_i=5.70$ \citep{Fukugita1995}, a constant that depends on the optical filter, and $m_{i}$ is the `modelMag' reported by the SDSS pipeline in Vega magnitudes. We have used a photon index $\Gamma = 1.8$ and a power-law spectrum. } &  $2.0^{4.7}_{-1.3}\times10^{\rm 43a}$ & $4.3\pm3.7\times10^{41}$\footnote{Assuming the [O {\sc{iii}}] emission  comes  from  an  optically  obscured  AGN,  we  estimate an upper limit for the intrinsic X-ray luminosity using the [O{\sc{iii}}]$\lambda5007$ luminosity (corrected for extinction) as a surrogate. Here, we adopt the mean calibration of \cite{Georgantopoulos2010} for obscured Seyfert-2, $\log(L_{\rm 2-10~keV}/L$[O {\sc{iii}}]$) = 1.16\pm0.77$. The $L_{2-10{\rm keV}}$ luminosity for source C has been converted to the 0.5-2 keV using a photon index of $\Gamma = 1.65$ \citep{Sazonov2008}.}\\
    
      &   &   &  \\

    Spectroscopy &  &  &  \\
    
Continuum slope &   $-0.66\pm0.03$  &   $-1.05\pm0.02$  &   $-1.30\pm0.05$  \\

[Ne {\sc V}]     &  [$112.7\pm13.6$, $1433.0\pm199.3$, $-625.8\pm106.3$]     &   \nodata    &    \nodata       \\
O{\sc{ii}}]$\lambda3727$     &  [$73.2\pm13.5$, $578.9\pm20.5$, $<$316.3]    &  [$245.9\pm30.7$, $1254.6\pm91.5$, $>$-19.7]       &          [$68.4\pm5.3$, $360.8\pm25.4$, $68.8\pm2.1$] \\
Ne {\sc{iii}}]$\lambda3869$ &  \nodata    &   [$28.9\pm2.8$, $1003.7\pm106.2$, $-162.5$]     &  \nodata     \\
Ne {\sc{iii}}]$\lambda3967$  &   \nodata &   [$18.3\pm2.4$, $772.9\pm111.3$, $-164.8\pm56.4$]      &\nodata   \\
H$\delta$ &  [$310.3\pm13.3$, $4795.0\pm237.9$, $>$-128.6]      &   [$26.8\pm2.3$, $597.3\pm53.0$, $-152.0\pm26.5$]      &      [$5.1\pm0.7$, $141.6\pm19.4$, $-0.3\pm0.1$] \\
H$\gamma$ &   [$700.4\pm31.0$, $4967.2\pm206.0$, $>$-47.0]      &    [$72.1\pm7.2$, $455.1\pm49.0$, $-138.5\pm16.2$]      &      [$13.2\pm0.1$, $190.7\pm16.6$, \\
            &   &   &  $-14.9\pm8.8$]  \\
H$\beta$  &  [$18.7\pm5.3$, $346.8\pm20.3$, $<$72.4]      &   [$149.2\pm4.4$, $581.6\pm19.3$, $-93.4\pm10.6$]     &   [$33.7\pm1.1$, $237.9\pm9.0$, $-23.2\pm4.8$] \\
H$\beta$(broad-1) &  [$310.2\pm55.9$, $1947.3\pm179.5$, $169.8\pm52.6$]      &      [$33.7\pm1.1$, $237.9\pm9.0$, $-23.2\pm4.8$]  &   \nodata    \\
H$\beta$(broad-2)      &  [$937.1\pm53.3$, $5083.0\pm243.3$, $360.7\pm90.0$]    &   \nodata     &    \nodata    \\
O {\sc{iii}}]$\lambda4959$   &  [$36.9\pm5.5$, $340.0\pm19.9$, $<$183.4]      &  [$26.1\pm5.2$, $520.2\pm34.1$, $>$-62.5]     &     [$12.1\pm1.4$, $335.6\pm43.2$, $8.7\pm2.6$] \\
O {\sc{iii}}]$\lambda4959$(broad) &  [$76.8\pm11.8$, $1602.2\pm121.3$, $<$42.7]    &  [$12.2\pm4.3$, $1362.7\pm200.2$, $>$-105.7]     &    \nodata    \\
O {\sc{iii}}]$\lambda5007$  &  [$102.8\pm5.2$, $336.7\pm19.7$, $<$203.4]      &   [$86.7\pm4.3$, $515.2\pm33.8$, $>$-42.5]     &     [$22.4\pm1.3$, $281.1\pm18.2$, $7.9\pm3.0$] \\
O {\sc{iii}}]$\lambda5007$(broad) &  [$136.3\pm12.2$, $1586.8\pm120.1$, $-591.2\pm108.0$]    &  [$25.1\pm6.8$, $1375.8\pm202.1$, $>$-59.6]      &  \nodata     \\
He {\sc i}$\lambda5876$    &  [$133.3\pm32.2$, $2814.4\pm333.9$, $-326.0\pm93.2$]      &  [$17.5\pm3.3$, $618.4\pm139.6$, $>$-69.7]      &    \nodata        \\
O {\sc{i}}]$\lambda6302$      & \nodata      &   [$73.2\pm3.1$, $754.0\pm37.1$, $-105.1\pm19.8$]     &         [$<$8.3, $<$128.2, $51.1\pm27.4$  ] \\
O {\sc{i}}]$\lambda6365$      &  \nodata     &  [$36.8\pm3.8$, $1108.3\pm132.8$, $>$-70.9]      &      [$<$8.3, $<$128.2, $51.1\pm27.4$] \\
N {\sc{i}}]$\lambda6529$      &   \nodata    &   [$20.9\pm2.5$, $543.7\pm60.3$, $119.7\pm28.5$]    &      \nodata       \\
N {\sc{i}}]$\lambda6548$       &  [$54.4\pm13.5$,$526.9\pm13.5$, $<$135.2]    &   [$56.6\pm14.2$,$523.9\pm29.0$, $>$-110.7]      &     [$16.4\pm2.0$, $234.7\pm35.7$, $-4.0\pm2.5$] \\
H$\alpha$      & [$160.9\pm13.5$, $524.1\pm25.9$, $<$294.3]     &  [$247.5\pm9.0$,$382.97\pm10.6$, $-77.7\pm4.9$]       &      [$121.2\pm1.9$, $204.4\pm3.5$, $-6.5\pm1.9$] \\
H$\alpha$(broad-1)      &  [$2545.3\pm54.7$, $2440.2\pm33.6$, $<$41.2]    &   [$747.8\pm25.4$, $1644.6\pm40.4$, $198.0\pm22.0$]     &    \nodata    \\
H$\alpha$(broad-2)      &  [$1473.2\pm47.4$, $9448.0\pm379.1$, $<$133.3]    &  \nodata      &   \nodata     \\
N {\sc{i}}]$\lambda6584$      &   [$318.3\pm20.2$, $525.7\pm20.7$, $<$110.6]   &  [$167.5\pm14.2$,$521.1\pm28.9$, $>$-42.6]     &          [$55.9\pm1.9$, $197.9\pm8.4$, $-8.3\pm4.5$] \\
S {\sc{ii}}]$\lambda6716$    &  [$26.1\pm7.0$, $262.7\pm63.7$, $-252.7\pm22.3$]      &  [$126.8\pm17.7$,$572.4\pm56.1$, $-91.4\pm33.1$]     &        [$27.95\pm6.22$, $186.33\pm61.78$, \\
      &   &   &  $-5.43\pm2.18$]  \\

S {\sc{ii}}]$\lambda6716$(broad)  &  \nodata   &  [$95.6\pm20.0$, $2479.9\pm505.4$,$<$195.2]    & \nodata  \\
S {\sc{ii}}]$\lambda6731$      &   \nodata    &   [$100.4\pm15.8$, $554.2\pm63.2$, $-108.3\pm39.9$]   &         [$17.1\pm4.8$, $227.3\pm63.6$, $2.7\pm3.1$] \\
    
    \tableline
    
    \label{table_full}
\end{tabular}
\end{table*}


\subsection{Spectral Fitting}

\begin{figure*}
    \includegraphics[trim=40 63 0 0, clip,scale=0.62]{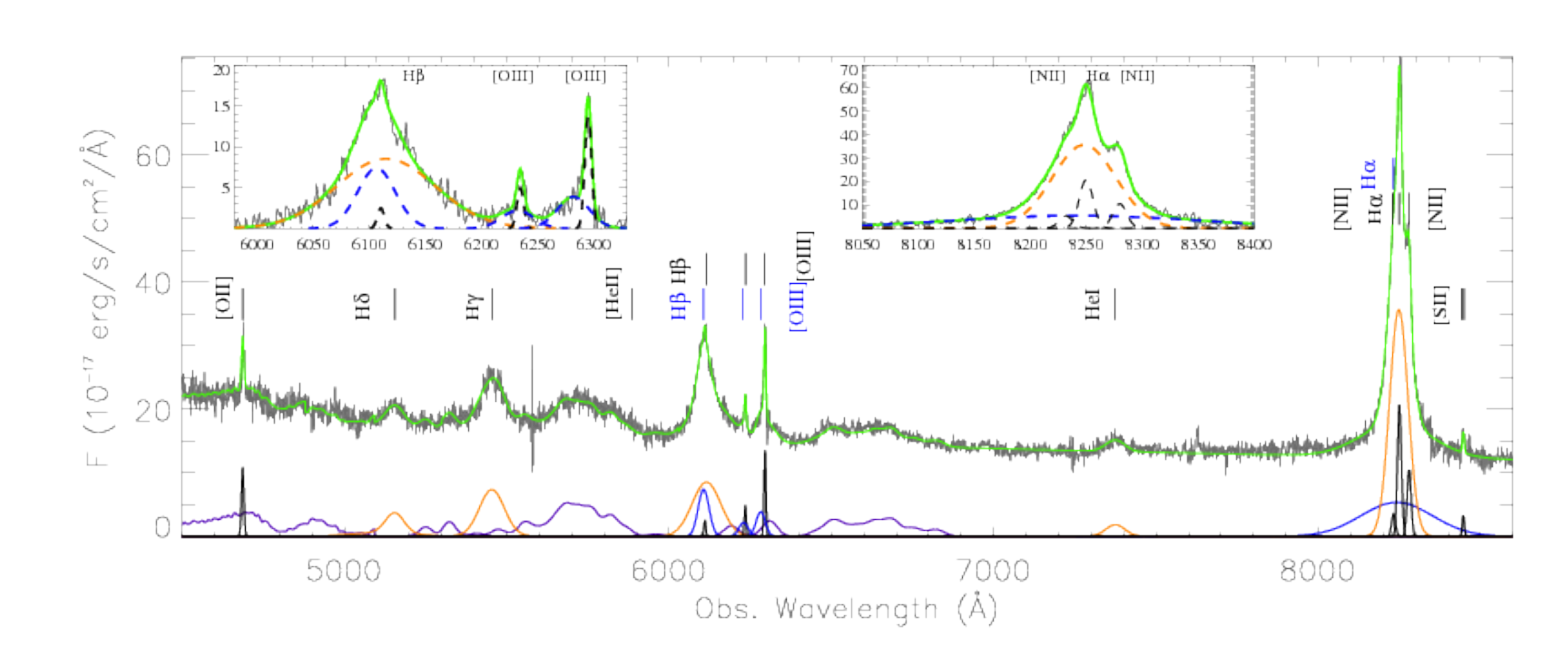}
    \hspace{-0.7cm}
    \includegraphics[trim=40 63 0 29, clip,scale=0.62]{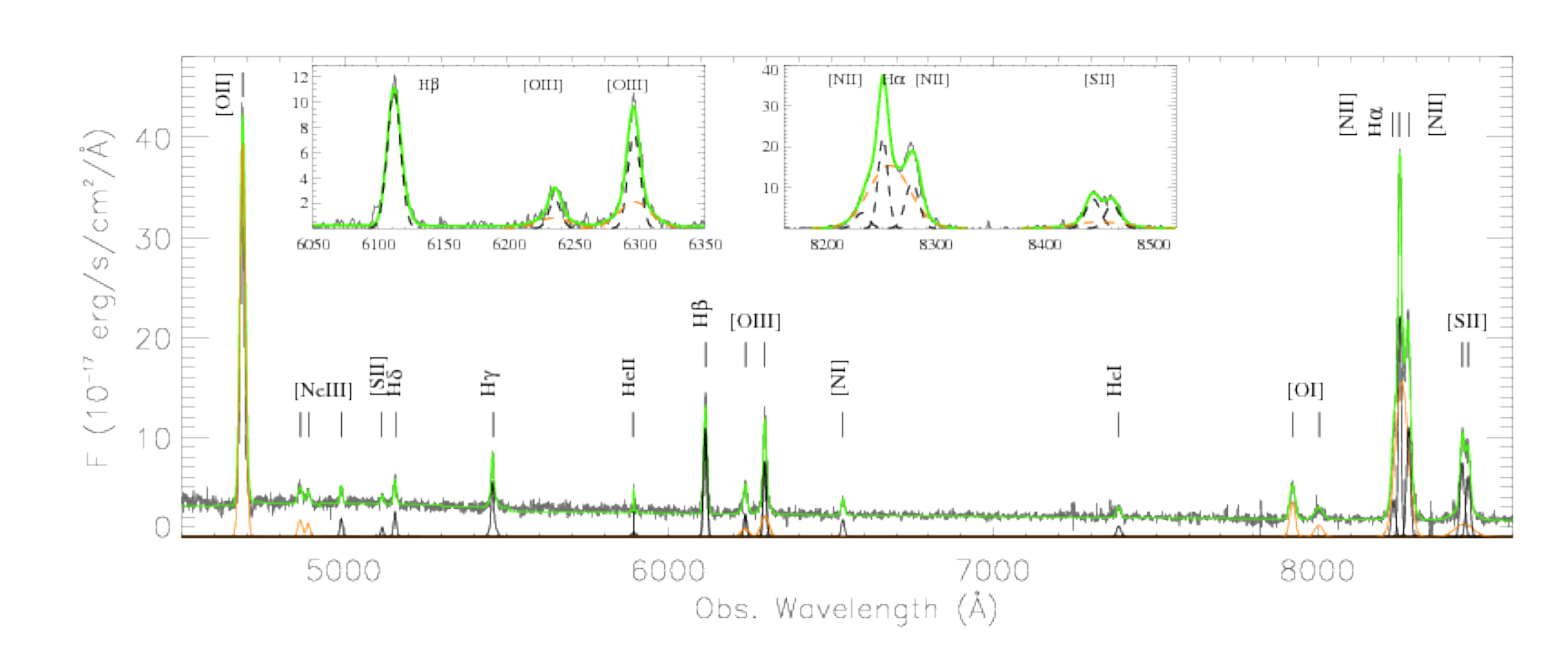}
    \includegraphics[trim=40 0 0 29, clip,scale=0.62]{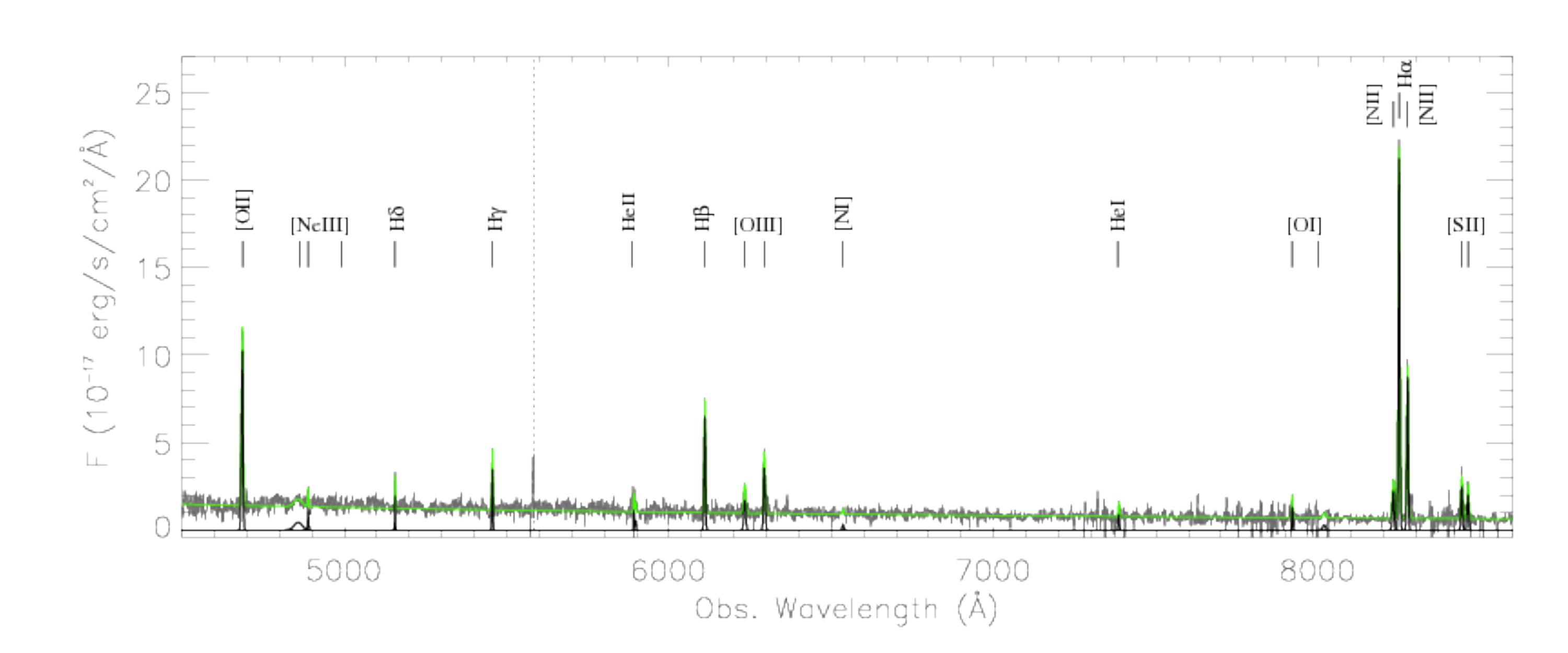}
    \caption{SDSS spectra of SDSS J1056+5516A, B and C nuclei from top to bottom. The gray and green lines represent the observed spectrum and the model sum, respectively. Orange: Set of broad emission lines (FWMH$>600~{\rm km~s^{-1}}$) at the redshift of the source. Black: Set of narrow emission lines at the redshift of the source. Blue: Set of blueshifted broad emission lines. Magenta: Fe {\sc{ii}} spectrum.}
    \label{fig:spectra}
\end{figure*}

Source A was observed by the original SDSS spectrograph \citep{Blanton2003}, while the sources B and C were observed by the BOSS spectrograph \citep{Dawson2013}\footnote{Further details can be found at the SDSS-III website.}. We examined the spectra of the objects to determine their optical properties using the IDL version of the mpfit fitting code \citep{Markwardt2009}. The SDSS spectra, corrected for the Galactic reddening of $E(B-V)=0.01$ mag, are displayed in Figure~\ref{fig:spectra} and the detected emission lines are listed in Table~\ref{table_full}.
    
We begin the fitting process by removing the effects of Galactic extinction using the \cite{Schlegel1998} map and a Milky Way extinction curve from \cite{Cardelli1989}, with $R_V= 3.1$. We identify the underlying continuum level using windows of 60~\AA~in featureless regions and we fit with a power law ($f_\nu \propto \nu^{\alpha}$). The Fe {\sc{ii}} multiplets are modeled by templates built from I Zw 1 ($\lambda>3500$~\AA, \citealp{Veron2004}; $\lambda<3500$~\AA, \citealp{Tsuzuki2006}). The redshifts of the Fe {\sc{ii}} lines are tied to the broad H$\beta$ component, while the normalization and the line-broadening are free parameters. The continuum and iron fit are then subtracted and the resulting spectrum is modeled by various functions. 

All the emission lines but H$\beta$, [O {\sc{iii}}]$\lambda \lambda 4959,5007$, and H$\alpha$ are fitted with a single Gaussian. For the H$\alpha$ line, we fit the observed wavelength range [8000,8500]~\AA. The narrow components of H$\alpha$ and [ N{\sc{ii}}]$\lambda \lambda 6548,6584$ are fitted with a single Gaussian and their velocity offsets and line widths are constrained to be the same. The relative flux ratio of the two [N {\sc{ii}}] components is fixed to 2.96. The broad H$\alpha$ component is modeled either with one or multiple broad Gaussians with FWHM~$>600~{\rm km~s^{-1}}$ \citep[e.g.,][]{Brescia2015}.
    
For the H$\beta$ line, we use the observed [5900,6400]~\AA~ wavelength range. Since the [O {\sc{iii}}]$\lambda \lambda 4959,5007$ lines frequently show an asymmetric blue wing, we model each of the narrow [O {\sc{iii}}] lines with two Gaussians, one for the core and the other for the wing. The flux ratio of the [O {\sc{iii}}] doublet is not fixed during the fit, but we found that the results show good agreement with the theoretical ratio of about 3. The velocity offset and FWHM of the narrow H$\beta$ line are tied to those of the core [O {\sc{iii}}] doublet components. The broad H$\beta$ line is modeled similarly to the broad H$\alpha$ one. The spectroscopic measurements with their $1\sigma$ errors are given in Table~\ref{table_full}.

\subsection{Emission-Line Diagnostics}

We used a combination of emission-line ratio (BPT diagnostics; \citealp[e.g.,][]{Baldwin1981}) and line widths to classify the sources in the system. Primarily, we classified narrow-line objects as having ${\rm FWHM_{H\alpha}<600~km~s^{-1}}$ and sources with ${\rm FWHM_{H\alpha}>600~km~s^{-1}}$ as potential type-1 AGNs \citep[e.g.,][]{Osterbrock1984,Brescia2015}.\footnote{We rely on the FWHM of the ${\rm H\alpha}$ line, rather than ${\rm H\beta}$, to ensure that we include intermediate-class AGNs (e.g., types 1.8,1.9, etc.)}

The nucleus A is a typical broad emission-line quasar. The spectrum of nucleus B shows a broad emission component of H$\alpha$ ($\sim 1645~{\rm km~s^{-1}} $) and relatively broad [O {\sc{ii}}] and [O {\sc{i}}] emission lines. Broad, but weak, components are also required to fit the [O {\sc{iii}}] emission line doublets. A broad component is also required for fitting the [S{\sc{ii}}] doublet. The $B$ nucleus satisfies the criteria we described above, and it is classified as an 1.8/1.9 AGN \citep{Osterbrock1984}.

BPT diagnostics suggest that C is a composite galaxy. Specifically, based on $\log($[N{ \sc{ii}}]/H$\alpha) = -0.34\pm0.01$ and $\log($[O {\sc{iii}}]/H$\beta) = -0.18\pm0.01$ ratios, $C$ is classified as a composite. C is classified as a star-forming galaxy based on the $\log($[S {\sc{ii}}]/H$\alpha) = -0.43\pm0.05$ ratio \citep{Kewley2006}. The [O {\sc{i}}] emission line is not detected at the $3\sigma$ level, therefore we can only estimate an upper limit of the ratio $\log($[O {\sc{i}}]/H$\alpha)<-1.17$, which places the source close to the boundary between star-forming and the low-ionization nuclear emission-line region galaxies (LINERs; \citealp{Heckman1980}). Thus, the classification of the nucleus C is somewhat ambiguous but with hints for hidden AGN activity \citep[e.g.,][]{Ho2008}.

\subsection{Black hole mass}
    
Assuming that the standard broad-line region scaling relations hold, we infer the BH mass of the quasar A from the monochromatic 5100~\AA~luminosity calculated from the power-law fit to the spectrum and the measured H$\beta$ line width of the broad component \citep{Vestergaard2006}, finding $M_{\rm BH,A}= 2.8\pm0.6\times 10^{8}~M_{\odot}$. As the source B does not show a broad H$\beta$ component, we estimate the BH mass from the broad H$\alpha$ component \citep{Greene2010}, where we find $M_{\rm BH,B}=2.4\pm0.3\times 10^{7}~M_{\odot}$.
    
For source C, we obtain a photometric spectral energy distribution (SED) using the SDSS broadband model magnitudes, corrected for extinction, and scaled to the $i$-band $c$-model magnitudes, the fitting code $HyperZ$ \citep{Bolzonella2000}, and the method and templates described by \cite{Maraston2013} to estimate its stellar mass. The SED-fitting outputs are given in Table~\ref{table_full}. From this estimated stellar mass, $M_{*C} = 1.5\pm0.4\times10^{10}~M_{\odot}$, we infer black hole masses using the local relation of \cite{Haring2004}, yielding a black hole mass of $M_{\rm BH,C} =2.4\pm0.9\times 10^{7}~ M_{\odot}$. Using the same relation, we have estimated the stellar masses of the systems A and B. The stellar mass ratio of the triple system A:B:C is about 7:1:1.

\vspace{2cm}

\section{Discussion}
    
The observations presented in this letter give new insights into the nature of this unique system. The differences in the spectra and colors of the three components suggest that we are dealing with a physical triple system and there is no possibility for a lens hypothesis. In addition, despite the observational evidence of clumpy \citep[e.g.,][]{Elmegreen2009} or dwarf galaxies \citep[e.g.,][]{Bournaud2007} in interacting systems, our analysis suggests that we are dealing with three distinct galaxies. While there is less double for the nature of components A and B, also the SED of C also points toward a typical galaxy, at least one order of magnitude more massive than the typical star-forming clumps and dwarf galaxies \citep[e.g.,][]{Elmegreen2009}, and without evidence for a young stellar population, $\sim3$~Gyr. Here, we investigate three different scenarios regarding the nature of these systems, each of which depends on the nature of the system's components.

\subsection{The General picture: An SMBH Triplet}

Here, we investigate the scenario that the nucleus C hosts an inactive SMBH, in contrast to A and B. In the general merger picture, if both galaxies involved in a merger host an SMBH, the formation of an SMBH binary is an inevitable stage of the process. Following the merger, the two SMBHs sink to the center of the system, forming a bound binary. When scaled to very massive binaries, the inferred lifetimes scales are of the order of a few Gyr. The long time span increases the possibility of a secondary merger before the final coalescence and the formation of an SMBH triplet \citep[e.g.,][]{Hoffman2007}. Based on recent studies, on average, all massive galaxies have experienced a merger in the last ten billion years \citep[e.g,][]{Bell2006}. Assuming uncorrelated events, and a typical binary lifetime of one billion years, then 10\% of SMBH binaries may form a triplet. Nevertheless, observational evidence of hierarchical triple SMBHs is limited.

The impact of a third SMBH can be crucial, either by ejection of one component \citep[e.g.,][]{Komossa2012}, by perturbing the orbit of the original binary system, or by supplying the binary with interstellar material even to the innermost regions, fueling the AGN/star formation activity. In the last case, the ongoing merger process may be related to the AGN activity in the system. It is therefore important to study the AGN activity of each component and associate it with the merging event. 
    
An important first clue comes from a comparison of the ROSAT X-ray measurement of the unresolved systems and the empirical calibration of the optical and [O {\sc{iii}}] emission to X-rays for the individual sources. We note that the archive X-ray and optical observations have a time separation of some decades and AGN sources are know to be highly variable. Despite the large uncertainties in our estimation, even the lower limit estimation for the combined X-ray emission of A, B and C is higher than that in the 2RXS ($L_{\rm 2RXS}\approx 3.0\times10^{43} < L_{\rm A+B+C,min }=3.9\times10^{43}~{\rm erg~s^{-1}}$ at 0.5-2 keV). The empirical estimation of the X-ray emission for the individual sources suggest a total enhancement of about $3.0^{+5.6}_{-1.7}$ times for the system.

If that is the case, it is possible that we observe the system at a high variability stage, or we are witnessing the ultimate fueling of the AGN activity in any of the three sources. The possibility of witnessing the onset of AGN activity in the nucleus C should be considered as an important event that could provide further insights into the role of mergers in the hierarchical growth of structures and AGN triggering. On the other hand, it is possible that the AGN activity in A and B is completely unrelated to the ongoing merger. If SMBHs grow their mass rapidly through a sequence of randomly oriented accretion events \citep[e.g.,][]{King2007}, it is possible that we are simply witnessing two galaxies hosting an AGN while are serendipitously interacting with a third one.

\subsection{Triple-AGN merger: A, B, and C harbor an AGN}

An SMBH triplet can become visible as a triple AGN when all three SMBHs accrete large amounts of gas at the same time. Therefore, AGN triplets are extremely rare, with only five candidates reported so far at separations $\lesssim10$~kpc (see \citealp{Deane2014}), the approximate effective radius for an elliptical galaxy. Due to their rarity, alternative explanations have been proposed, such as growing via merger-induced accretion, or growing in situ from seed BHs that collapsed within the preceding few hundred million years \citep[e.g.,][]{Schawinski2011}. 
    
While the optical spectra of the three nuclei suggest that they all potentially host AGNs, optical identification alone is inconclusive, especially for source C. The source of ionization in transition nuclei, such as source C, has been the subject of much debate. Summarizing results from X-ray surveys of nearby galaxies, \cite{Ho2008} concluded that the majority of transition nuclei do contain AGNs. Thus, the nucleus C can be interpreted as possibly containing low-luminosity accretion-powered AGN. Deeper X-ray and/or radio observations can help to pin down the nature of the system.
    
Indeed, although limited observational constraints exist on triple-AGN systems, their fraction of associated radio emission is significantly higher ($>60$\% per cent; \citealp{Deane2014}) than that for single AGNs ($\sim 10$\%; e.g., \citealp{Ivezic2002}). This fact might suggest that triple systems lead to higher accretion activity and consequently a higher chance of jet triggering. In this scenario, the radio emission associated with the nucleus B is consistent with hydrodynamical simulations that find peak accretion occurs at small separations \citep[e.g.,][]{VanWassenhove2012}.
    
Among the five $\lesssim10$~kpc candidate triple-AGN systems, it is the only one with clear merger signatures, detected in both X-ray and radio. If the triple-AGN nature is confirmed, it will offer a unique opportunity to extend the limited observation constraints that exist on the evolution, feeding, and feedback processes of these systems. Due to its separation, all the system's components can be easily resolved by available facilities, providing an excellent target of opportunity for multi-wavelength monitoring campaigns. 
    
\subsection{Ejection Hypothesis: A quasar, a nearly naked SMBH, and a remnant host galaxy}
    
Two ejection mechanisms have been suggested for the displacement of an SMBH from its host galaxy: gravitational radiation recoil during the coalescence of a binary SMBH \citep[e.g.,][]{Komossa2012}; or a gravitational slingshot involving three SMBHs \citep[e.g.,][]{Hoffman2007}. Gravitational waves from the coalescence of SMBHs are believed to generate kick velocities up to several thousand ${\rm km~s^{-1}}$ on the remnant SMBH \citep[e.g.,][]{Komossa2012} while, the intrusion of a third galaxy may favor the displacement \citep[e.g.,][]{Hoffman2007}. 
    
An observable recoiling SMBH can be distinguished from a stationary one via offsets in either physical or velocity space. A displacement of $\lesssim20$~kpc from the nucleus, as observed in the cases of B and C, is in agreement with the predictions from hydrodynamic/$N$-body simulations \citep[e.g.,][]{Blecha2011}. In this case, the SMBH is statistically more likely to have a relatively low velocity, since a large fraction of orbital time is spent at turnaround \citep[e.g.,][]{Blecha2008}. Likewise, the largest velocity offsets ($\upsilon_{\rm LOS, B-C}=170~{\rm km~s^{-1}}$) will occur either soon after the recoil event or on subsequent passages of the SMBH through the galactic disk.
    
In this hypothesis, the ${\rm B+C}$ system underwent a merger with quasar A, as the strong tail suggests. Upon merging, the B SMBH recoiled, removing part of the the broad-line region and high ionization gas that bound to the SMBH, leaving behind the narrow-line region (C). As the B SMBH is still observable, it must carry an accretion disk, giving a possible explanation for the exceptionally broad Neon emission lines \citep{Komossa2008}, and be detected during a high-accretion stage. \cite{Blecha2011} have found an ejected disk mass peak at $M_{\rm disk,ej}\sim 0.1M_{\rm SMBH}$ for random, dry spin mergers, while the typical disk masses are slightly higher ($\sim 0.03-0.3M_{\rm SMBH}$) in hybrid and aligned models. Assuming that the ejected disk feeds the SMBH at a rate $\dot M_{\alpha}$ and the SMBH accretes only from the direction of its disk plane, we can estimate the accretion rate that is converted into the bolometric luminosity $\dot M_{\alpha, B}=L_{\rm bol, B}/\epsilon c^{2} \approx 0.05~M_{\odot}~{\rm yr^{-1}}$ \footnote{The bolometric luminosity for source B is computed using the 5100~\AA~luminosity and a bolometric correction of $7.79\pm1.69$ \citep{Krawczyk2013}; $L_{\rm bol,B} = 2.7\pm0.6\times10^{44}~{\rm erg~s^{-1}}$,}, where $\epsilon_{\rm rad}=0.1$ is the radiative efficiency at high Eddington ratios \citep[e.g.,][]{Marconi2004}. 
    
If the SMBH is out of the plane of the disk, its accretion rate is ̇$\dot M_{\alpha}$ for $0<t<t_{\rm accr}$, where $t_{\rm accr, B}= M_{\rm disk,ej}/\dot M_{\rm \alpha, B}\approx 48~{\rm Myr}$ is the accretion time-scale due to the ejected disk \citep[e.g.,][]{Blecha2008}. During the disk lifetime, the ejected SMBH has traversed a projected distance of $d_{\rm proj, B-C} \lesssim 20~{\rm kpc}$. Thus, we can estimate the projected ejection velocity $\upsilon_{\rm proj,ej} \approx 410~{\rm km~s^{-1}}$. Our results are in perfect agreement with \cite{Blecha2008} predictions for the type of recoiling SMBH of our system ($M_{\rm BH}=10^6 - 10^8~M_{\odot}$; $\upsilon_{\rm ej} = 400-1000~{\rm km~s^{-1}}$).
    
If the gravitational-wave recoil hypothesis is confirmed, the observed system will be one of the best examples supporting recent recoiling black holes simulations that argue that spatially offset AGNs are naturally more longer-lived than kinematically offset AGNs \citep[e.g.,][]{Blecha2011} and thus easier to detect.

\section{Conclusions}
    
The SDSS J1056+5516 system is incredibly rich in interesting phenomena. The observations and the analysis of this triple system presented in this Letter give new insights into its peculiar nature. The information gathered here points to three possible scenarios, each of which is very important to our understanding of the hierarchical growth of galaxies, AGN triggering, or even-gravitational waves production. 
    
The triple system clearly warrants further investigation. X-ray and radio observatiosn should rapidly narrow the possible origins of this intriguing system. Another point of interest is to look for further evidence of gas morphology and kinematics in order to investigate the ejection hypothesis, by mapping the cold and warm molecular gas through CO and H$_{2}$ emission, as well as more ionized gas through [O {\sc{iii}}] and H$\alpha$ emission.

\acknowledgments

We thank Marios Karouzos for helpful discussions and comments. We thank Pedro Rodriguez for his help with the combined SDSS image presented in this Letter.

\end{document}